\documentclass[10pt]{ismd08}
\usepackage{graphicx}
\usepackage{cite,./mcite}
\usepackage[latin1]{inputenc}

\newcommand{\si}[0]{\ensuremath{\sigma}}

\newcommand{\abs}[1]{\ensuremath{\left| #1 \right|}}

\newcommand{\ga}[0]{\ensuremath{\gamma}}

\newcommand{\La}[0]{\ensuremath{\Lambda}}

\newcommand{\al}[0]{\ensuremath{\alpha}}

\newcommand{\bl}[1]{\mathbf{#1}}

\newcommand{\rmax}[0]{\ensuremath{r_{\max}}}
\newcommand{\text}[1]{\mathrm{#1}}
\begin{document}
\title{Quasielastic Scattering in the Dipole Model\footnote{Work supported in part by the Marie
    Curie RTN ``MCnet'' (contract number MRTN-CT-2006-035606).}}
\author{Christoffer Flensburg\protect\footnote{\ \ In collaboration with G\"osta Gustafson and Leif L\"onnblad}}
\institute{Dept.~of Theoretical Physics,
  S\"olvegatan 14A, S-223 62  Lund, Sweden}
\maketitle
\vspace {-2.45cm}
\begin{flushright}
LU-TP 09-03 \\
MCnet/09/02
\end{flushright}
\vspace {0.6cm}
\begin{abstract}
A series of previous papers \cite{Avsar:2005iz,*Avsar:2006jy,*Avsar:2007xg} develops a dipole model in initial state impact parameter space that includes subleading effects such as running $\al_s$, unitarity, confinement and saturation. Here some recent work \cite{Flensburg:2008ag} is presented, where the model is applied to a new set of data: vector meson production in $\ga^\star p$, DVCS and $d\si /dt$ in $pp$. This allows us to tune a more realistic model of the proton wavefunction from the $pp$ data, and confirm the predictive power of the model in high $Q^2$ of DVCS and vector meson production. For low $Q^2$ vector meson resonances dominate the photon wavefunction, making our predictions depend on a tuned parametrisation in this range.
\end{abstract}
\vspace {-0.4cm}

\section{Why Dipoles?}
\label{sec:why}

To calculate cross sections for hadronic particles it is important to understand the evolution in the initial state. In a high energy collision, each of the two incoming particles will emit gluons before meeting and interacting. Enumerate the possible initial states with $i,j$ and give each state a probability $w_i$ such that $\sum_i w_i = 1$. With a scattering probability $p_{ij}$ between state $i$ and $j$ the total interaction probability can be expressed as
\begin{equation}
T_{\text{tot}}(\bl{b}) = 2 \sum_{ij} w_i w_j p_{ij} .
\end{equation}
That means that the expectation value of $p_{ij}$, weighted by $w_i$ can be measured. Similarly the diffractive, including elastic, cross section is
\begin{equation}
T_{\text{diff}}(\bl{b}) = \sum_{ij} w_i w_j p_{ij}^2 .
\end{equation}
To get both these cross sections right, not only the expectation value of $p_{ij}$ with respect to $w_i$ is required, but also the fluctuations. That is, it is possible to measure if the cross section is dominated by frequently occuring states with a low interaction probability, giving a low $T_{\text{diff}}/T_{\text{tot}}$, or by rare states with a high interaction probability, giving a high $T_{\text{diff}}/T_{\text{tot}}$. Also the elastic interaction probability can be written in this way as
\begin{equation}
T_{\text{el}}(\bl{b}) = \left( \sum_{ij} w_i w_j p_{ij} \right)^2 .
\end{equation}
This makes the form of the impact parameter profile important since the more spread out the interaction probability is, the smaller the elastic cross section will be.

These arguments show that to describe all the above cross sections, it is important to have a good description of the fluctuations, both in $\bl{b}$ and $w_i$.

\section{Our Model}
\label{sec:model}
Our model uses colour dipoles in impact parameter space, based on the model by Mueller \cite{Mueller:1993rr,*Mueller:1994jq,*Mueller:1994gb}. One of the reasons to do the calculations in impact parameter space is that each emission is on a shorter timescale than the previous ones, essentially freezing their transverse position. Each incoming particle is represented by a dipole state (for example the photon is represented as a single dipole), which is then evolved in rapidity before colliding. The evolution is equivalent to leading order BFKL, and we have made corrections for higher order effects.

\subsection{Evolution}
Each dipole is emitting gluons, forming two new dipoles with a probability density of
\begin{equation}
\frac{d\mathcal{P}}{dY} = \frac{\bar{\alpha}(r_<)}{2\pi \rmax^2} d^2z \left( \frac{\bl{x}-\bl{z}}{\abs{\bl{x}-\bl{z}}} K_1(\frac{\abs{\bl{x}-\bl{z}}}{\rmax}) - \frac{\bl{y}-\bl{z}}{\abs{\bl{y}-\bl{z}}} K_1(\frac{\abs{\bl{y}-\bl{z}}}{\rmax}) \right)^2 \label{eq:emprob}
\end{equation}
where $\bl{x}$ and $\bl{y}$ are the transverse positions of the partons in the original dipole, while $\bl{z}$ is the position of the emitted gluon. $r_<$ is the size of the smallest of the three involved dipoles (the original one, and the two new ones), and is setting the scale for the \textbf{running coupling constant} for the emission. Also \textbf{confinement} is included in this emission density, which takes form in the modified Bessel functions $K_1$ which fall off exponentially for large arguments. The confinement scale is set by $\rmax$, corresponding to a gluon mass $1/\rmax$ in a screened Yukawa potential.

\textbf{Energy conservation} is accounted for by approximating the $\bl{p_T}$ of the partons as twice the inverse dipole size, from which $p_+$ can be calculated. Allowing only emissions that respect energy-momentum conservation gives a cutoff for emitting too small dipoles, that is, too large $\bl{p_T}$, cutting away the poles in the emission probability (\ref{eq:emprob}).

Apart from the 1 to 2 emission above, the model also includes a 2 to 2 dipole swing, where dipoles of the same colour may recombine, changing the colour flow, but not the momenta. The swing favours small dipoles over large dipoles, which reduces the cross section and gives a \textbf{saturation} effect.

\subsection{Interaction and Cross sections}
To find the cross section, the interaction probability of two evolved states of dipoles is calculated for a given impact parameter. The probability that a dipole $i$ from one state will interact with a dipole $j$ in the other state is
\begin{equation}
f_{ij} = \frac{\al_s^2}{8} \left( \log \left( \frac{(\bl{x}_i-\bl{y}_j)^2(\bl{y}_i-\bl{x}_j)^2}{(\bl{x}_i-\bl{x}_j)^2(\bl{y}_i-\bl{y}_j)^2} \right) \right)^2 ,
\end{equation}
with $\bl{x}_i$, $\bl{y}_i$ the transverse positions of the partons of dipole $i$. This is then corrected for confinement, which introduces Bessel functions as was done for the emission probability (\ref{eq:emprob}). Using this, the total interaction probability of the two dipole states can be calculated in the \textbf{unitarised} form,
\begin{equation}
T(b) = 1-e^{-\sum f_{ij}} .
\end{equation}
This is again using the fact that the interactions are taking place during a short timescale, freezing the transverse positions of the partons. This evolution and interaction can be simulated in a Monte Carlo program to determine the interaction probability numerically. Integrating over the impact parameter then gives the total cross section, and modifications to the order of integration as in section \ref{sec:why} yields diffractive and elastic cross sections.

\section{Results}
\begin{figure}
  \includegraphics[angle=0, width=0.5\linewidth]{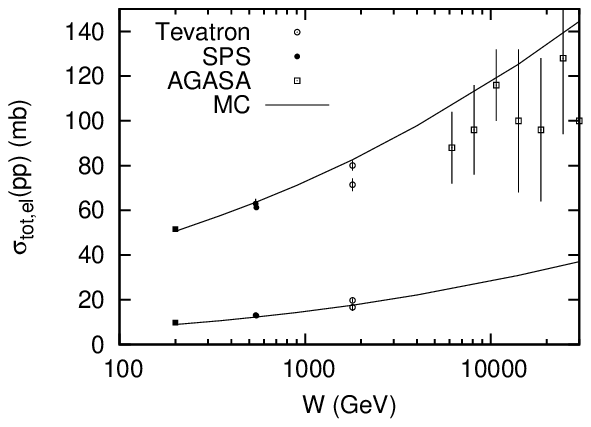}%
  \includegraphics[angle=0, width=0.5\linewidth, height=0.35\linewidth]{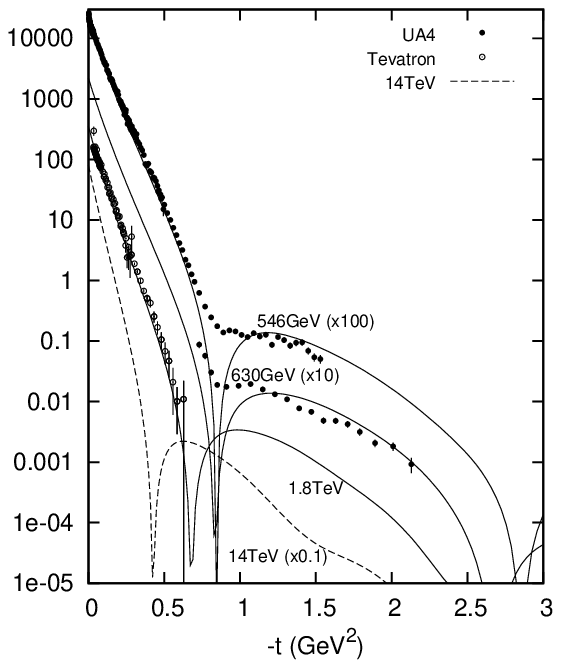}
\caption{\label{fig:pp} Left: The total and elastic $pp$ cross section. Right: Differential $d\si /dt$ cross section in $pp$. Data from \protect\cite{ Abe:1993xx }. }
\end{figure}

By tuning the two evolution parameters $\La_{\text{QCD}}$ and $\rmax$ and the proton wavefunction we can describe the total and elastic $pp$ cross section (fig \ref{fig:pp}). The tuned proton wavefunction is an equilateral triangle of dipoles with a radius of 3~GeV$^{-1}$. It should be noted that once the cross section is tuned for a total and elastic cross section at a given energy, the energy dependence of the cross sections depends very weakly on the tuning, so it is a direct result of the evolution in our model. The fourier transform of the elastic amplitude then gives also $\si (t)$. As the elastic amplitude is calculated through the optical theorem, only the imaginary part is included, which causes a dip to 0 amplitude at a certain $t$. With the real part included, this dip would be smoothed out. The fact that it is possible to describe the energy dependence of the cross sections, as well as following $\si (t)$ over many orders of magnitude is a sign of the predictive power of the model.

\begin{figure}
  \includegraphics[angle=0, width=0.5\linewidth, height=0.27\linewidth]{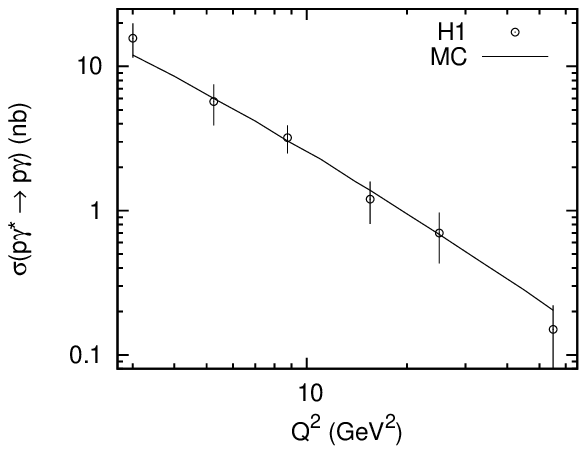}%
  \includegraphics[angle=0, width=0.5\linewidth, height=0.27\linewidth]{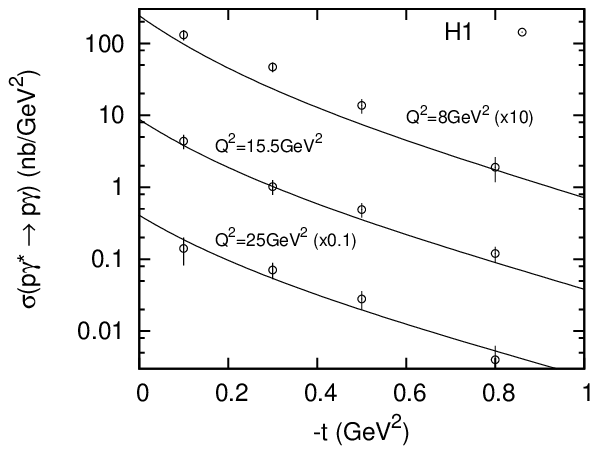}
\caption{\label{fig:pg} Top: Total $\ga^\star p$ as function of $Q^2$ (left) and $W$ (right). Bottom: DVCS for $W=82$~GeV as function of $Q^2$ (left) and $t$ (right). Data from \protect\cite{Chekanov:2005vv,:2007cz} }
\end{figure}

It is possible to calculate also $\ga^\star p$ using the virtual photon dipole wavefunction. For high $Q^2$ the wavefunction can be calculated perturbatively and the cross section as function of $Q^2$ and $W$ is predicted directly from the $pp$ tuning. The results agree with data (dotted line in fig \ref{fig:pg}), showing that the model can predict data without being tuned to it.

\begin{figure}
  \includegraphics[angle=0, width=0.5\linewidth, height=0.28\linewidth]{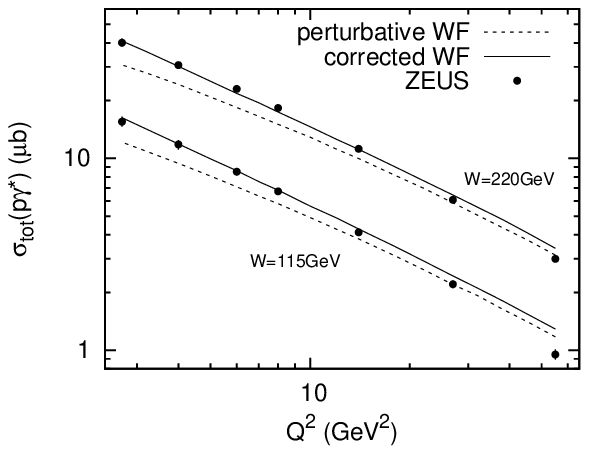}%
  \includegraphics[angle=0, width=0.5\linewidth, height=0.28\linewidth]{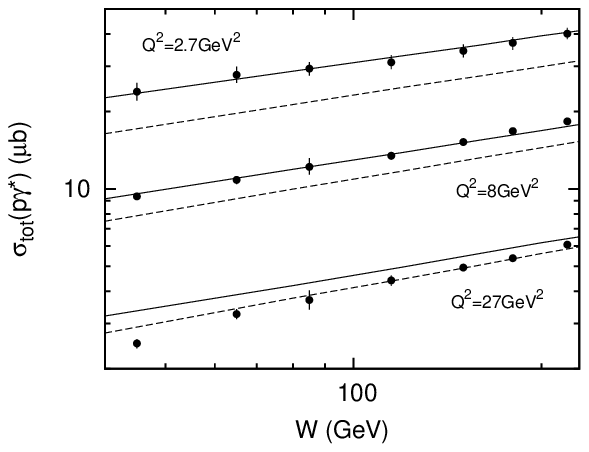}
\end{figure}

For low $Q^2$ (below 5-10~GeV$^2$) the photon wavefunction will have important soft contributions. Confinement suppresses too large dipoles, which can be taken into account by shrinking large dipoles coming out from the perturbative wavefunction. This can be compared to the confinement used in the evolution and can be estimated using the confinement scale $\rmax$ from the evolution. The most important effect is when the quark-antiquark pair propagates as a vector meson, boosting the wavefunction at mesonic dipole sizes. This vector meson resonance is not well understood quantitatively, so it had to be parametrised and tuned to low $Q^2$ total $\ga^\star p$ cross section data. The result with both soft effects included in the photon wavefunction is shown in the full line in fig \ref{fig:pg}.

Once the photon wavefunction was determined, also for low $Q^2$, the deeply virtual Compton scattering (DVCS) cross section can be calculated, using a $Q^2=0$ photon wavefunction for the outgoing particle. The results agree with data in $Q^2$, $W$ and $t$ dependence as can be seen in the plots in fig \ref{fig:pg}, further confirming the predictive power of our model.

\begin{figure}
  \includegraphics[angle=0, width=0.5\linewidth, height=0.28\linewidth]{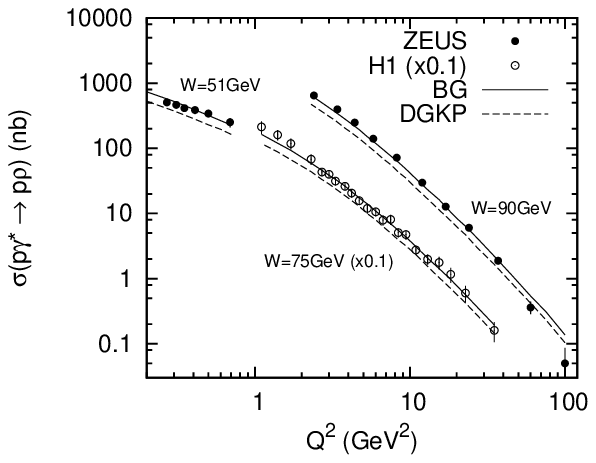}%
  \includegraphics[angle=0, width=0.5\linewidth, height=0.28\linewidth]{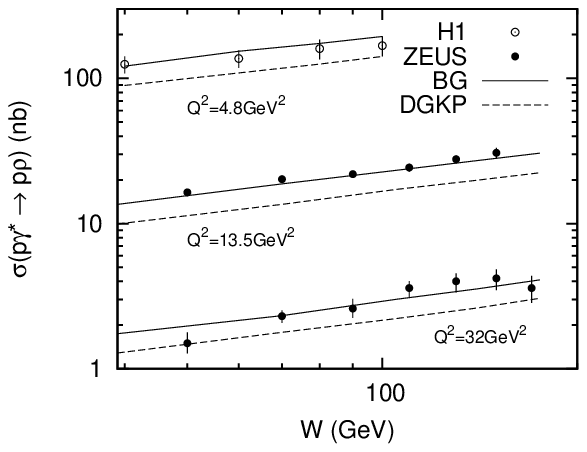}
\caption{\label{fig:rho} Rho production as function of $Q^2$ (left) and $W$ (right). Data from \protect\cite{Adloff:1999kg,Chekanov:2007zr}}
\end{figure}

By replacing the outgoing $Q^2=0$ photon wavefunction with a vector meson wavefunction, we can also calculate vector meson production cross sections. The vector meson wavefunction cannot be calculated perturbatively, but there are several models that estimate it, using normalisation and decay width to fix parametrisations. We used the DGKP \cite{Dosch:1996ss} and the Boosted Gaussian \cite{Forshaw:2003ki} models in our calculations. For the light vector mesons, the $Q^2$ and $W$ dependence on the total cross section agrees well with data, specially for the Boosted Gaussian model (fig \ref{fig:rho}). Also the $t$ dependence agrees for high $Q^2$, while for lower $Q^2$, the slope is too steep. This is not surprising, as the vector meson dominance of the photon wavefunction dominates in this range. It was tuned only to the total cross section in $\ga^\star p$, and we can not expect this parametrisation to correctly describe also the impact parameter profile that determines the $t$ dependence. Possibly, this is also the case in DVCS, but since the available experimental data for $t$ dependence does not go below Q$^2=$ 8~GeV$^2$, it is not observed. Moreover, the vector meson wavefunctions are approximative parametrisations, and they may yield incorrect $t$ distributions.

Also $\psi$ production can be calculated with this method, however, the results are not as good. One source of uncertainty is the vector meson resonance correction to the photon wavefunction, which would have to be retuned for a charm pair fluctuating into a $\psi$. More work is needed to achieve reliable results for heavy quark vector mesons.

\section{Conclusions and outlook}
Our dipole model has proven to describe a wide selection of data in both $pp$ and in $\ga^\star p$ collisions. The $pp$ data and the total $\ga^\star p$ cross section has been used for tuning the parameters of the evolution and the wavefunctions, while other aspects, like DVCS and the energy dependence of all processes, have been found without tuning, showing good predictive power of the model. For low $Q^2$ there are soft effects in the photon wave functions that we do not understand quantitatively, mainly the vector meson resonance.

Looking forward, we are currently working on using the information in the evolved states to determine not only the cross section, but also the exclusive final state. The evolution gives us the particles, their momenta, and even their colour connections. Some of the partons that have not collided will, however, have spacelike momenta and have to be reabsorbed as virtual fluctuations.

%
%
%

\begin{footnotesize}
\bibliographystyle{ismd08} 
{\raggedright
\bibliography{ismd08}
}
\end{footnotesize}
\end{document}